# NEXT-GENERATION CYBERATTACK DETECTION WITH LARGE LANGUAGE MODELS: ANOMALY ANALYSIS ACROSS HETEROGENEOUS LOGS

Yassine Chagna[1] and Antal Goldschmidt [2]

[1,2] School of Engineering, Coventry University, Coventry, UK

[1]chagnay@coventry.ac.uk, [2] ab2216@coventry.ac.uk

## ABSTRACT

*This project explores large language models (LLMs) for anomaly detection across heterogeneous log sources. Traditional intrusion detection systems suffer from high false-positive rates, semantic blindness, and data scarcity logs are inherently sensitive, making clean datasets rare. We address this through: (1) LogAtlas-Foundation-Sessions and LogAtlas-Defense-Set balanced, heterogeneous log datasets with explicit attack annotations and privacy preservation; (2) empirical benchmarking revealing why standard metrics (F1, accuracy) mislead security applications; (3) a two-phase training framework combining log understanding (Base-AMAN, 3B parameters) with real-time detection (AMAN, 0.5B parameters via knowledge distillation). Results demonstrate practical feasibility: 0.3–0.5s inference per session, <$50/day operational cost.*

## 1. Introduction

Log data provides a rich record of system, network, and application activity, yet its scale and diversity make anomaly detection fundamentally challenging. Traditional rule-based and machine learning approaches fail to generalize across heterogeneous log sources, leading to high false-positive rates that overwhelm incident detection and response teams. These conventional methods struggle with two core limitations: they cannot capture semantic information embedded in naturally written log messages, and they require extensive labeled datasets for training, which are rarely available in sufficient quantities for the minority class of malicious activities.

Large language models have emerged as a promising direction due to their ability to parse, summarize, and semantically interpret large volumes of text-like data. Recent LLM-based systems (LogLLM, Audit-LLM, LogRESP-Agent) demonstrate that LLMs can enhance anomaly detection, provide contextual explanations, and reduce noise. However, existing systems face critical limitations: they are often limited to single log types, tested on restricted datasets, or impractical for deployment due to computational cost. Furthermore, applying LLMs to log analysis introduces specific challenges including context length limitations, the requirement that semantic understanding extend beyond individual entries to capture sequential and behavioral patterns, and the need for outputs that are accurate, interpretable, and actionable for security analysts.

This project addresses these challenges through a two-phase training strategy: (1) Base-AMAN learns general log understanding from heterogeneous sources without anomaly-detection-specific bias, capturing syntax, templates, and temporal structures; (2) AMAN distills this into a lightweight model for real-time detection. This two-phase design preserves semantic depth while remaining computationally feasible for deployment, moving beyond simple binary classification toward an adaptive intrusion detection paradigm that reduces false alarms, highlights risky anomalies, and supports root cause analysis.

This paper is structured as follows: Section 2 examines the state-of-the-art in log analysis and research gaps; Section 3 details dataset construction with explicit attack annotations and balanced class distributions; Section 4 demonstrates how data distribution impacts model performance; Section 5 presents the two-phase training strategy; Section 6 covers results and practical contributions; Section 7 concludes with implications and future directions.

## 2. Related Works

The study of log anomaly detection has evolved rapidly with the advent of large language models and advanced training paradigms. This section examines the state-of-the-art in log-based anomaly detection, focusing on existing solutions, training approaches, and research gaps.

### 2.1 Data Scarcity and Log Dataset Challenges

Log anomaly detection presents significant data acquisition challenges distinct from other AI domains. Logs frequently contain sensitive personal and professional information, creating legal and ethical barriers that prevent organizations from openly sharing data. Even when logs are available, they typically exist as raw, unannotated records with inconsistent schemas, hindering reproducible benchmarking. Loghub has emerged as the primary repository, aggregating 19 distinct log datasets totaling over 77 GB from distributed systems, supercomputers, operating systems, and mobile platforms. However, these datasets are now 3–5 years old and lack representation of modern cloud-native architectures and emerging attack vectors. The need for expanded, high-quality open resources with carefully curated attack annotations and privacy-preserving transformations motivates ongoing efforts to develop next-generation benchmarks.

*Table 1 Key Public Log Datasets*

| Dataset | Unique Log Keys | Log Sequences | Avg Sequence Length | Normal | Anomalous | Anomalous Distribution % |
|---|---|---|---|---|---|---|
| HDFS | 48 (15) | 575,061 | 19 | 553,223 | 16,838 | 2.95 |
| BGL | 396 (160) | 36,927 | 58 | 28,631 | 3,296 | 10.32 |
| Thunderbird | 7,703 (904) | 112,959 | 166 | 67,039 | 40,920 | 37.90 |

HDFS, BGL, and Thunderbird represent the most widely utilized benchmarks in LLM-based log anomaly detection. Despite widespread adoption, these datasets are now several years old. The development of new, high-quality datasets with improved annotation quality, diverse attack scenarios, and balanced class distributions has become increasingly crucial.

### 2.2 Fine-Tuning and Semantic Log Representation

Fine-tuning pretrained language models on domain-specific log data has emerged as the most empirically successful approach for log anomaly detection. Key approaches include self-supervised learning with masked language modeling (LogBERT), parser-free methods using token-level prediction (LAnoBERT) or direct semantic extraction (NeuralLog), and integrated architectures combining BERT with larger language models (LogLLM). All achieve F1-scores exceeding 0.95 on benchmark datasets, with principal trade-offs between parsing requirements, inference latency, and handling of template evolution.

The parser-free approaches address a critical practical limitation: explicit parsing introduces errors on out-of-vocabulary entries and semantic misunderstandings in heterogeneous environments. Integrated LLM architectures like LogLLM add robustness to template evolution of a real-world challenge as systems are updated. These fine-tuning approaches share a core principle: log sequences exhibit intrinsic statistical regularities where normal operations follow predictable patterns and anomalies manifest as deviations, which fine-tuning teaches models through self-supervised learning.

### 2.3 Research Gaps and Future Directions

Despite substantial progress in LLM-based log anomaly detection, significant technical gaps persist. Root cause analysis remains severely underdeveloped: most systems output only binary anomaly scores or confidence metrics, providing security analysts with minimal actionable insight. Computational efficiency at inference time remains a critical barrier: most models require 1–5 seconds per log session, while real-time security operations demand sub-100-millisecond latency. Knowledge distillation presents a promising approach to address this challenge by reducing model size while maintaining detection accuracy. Cross-domain generalization remains inadequately addressed: nearly all published models achieve strong performance on single log types or single organizational contexts, yet fail when deployed across new log sources. Interpretability and transparency require substantial advancement: most LLM-based approaches suffer from the "black box" problem where explaining model decisions remains challenging.

*Table 2 Fine-Tuned LLMs for Log Anomaly Detection*

| System | Base Model | Fine-Tuning Strategy | Parsing Requirement | Datasets Evaluated | F1 Score | Key Advantage | Limitation |
|---|---|---|---|---|---|---|---|
| **LogBERT** | BERT | Masked LM + Hypersphere | Yes | BGL, HDFS, Thunderbird | 0.95+ | Detects deviations well | Template-specific |
| **LAnoBERT** | BERT | Token-level prediction | No | BGL, HDFS, Thunderbird | 0.95+ | Parser-free | Computational cost |
| **NeuralLog** | BERT | Semantic vector extraction | No | BGL, HDFS | 0.96 | Context-aware | Resource intensive |
| **LogLLM** | BERT + Llama | Three-stage training | No | BGL, HDFS, Thunderbird, HPC | 0.97+ | Handles evolving templates | High computational cost |

## 3. Data Preparation and Dataset Design

### 3.1 Foundation Dataset: AIT Log Data Set (AIT-LDS v2.0)

The primary experimental corpus is the AIT Log Data Set version 2.0, a collection of synthetic multisource logs from eight enterprise testbeds with line-level ground truth annotations. Each testbed simulates a realistic environment containing mail servers, file shares, WordPress servers, VPN services, firewalls, and monitoring hosts; normal behavior is generated continuously with attack steps injected at known times. The AIT collection provides three critical properties: diverse log types (Apache, authentication, DNS, VPN, Suricata alerts, syslog, audit logs, network packet captures), explicit attack labels, and metadata describing host inventories and simulation time windows.

### 3.2 Derived Datasets: LogAtlas-Foundation-Sessions and LogAtlas-Defense-Set

Building upon AIT v2.0, we engineered two carefully constructed datasets optimized for distinct training phases. These datasets, publicly available on Hugging Face as LogAtlas-Foundation-Sessions and LogAtlas-Defense-Set, serve fundamentally different purposes: the first phase prioritizes broad, general-domain log understanding across diverse systems without anomaly-detection-specific bias, while the second phase focuses on developing discriminative detection capabilities through exposure to balanced attack and normal examples.

**LogAtlas-Foundation-Sessions (Pretraining Dataset):** Consists of over 44,000 temporal sessions aggregating more than 19 million raw log events. Sessions are constructed by grouping consecutive log entries sharing contextual attributes (host, process, user) with temporal boundaries where gaps of five minutes or longer indicate distinct operational contexts. The dataset preserves the natural class distribution (~2% attack prevalence), a critical design choice for general log understanding. Each session is annotated with rich metadata: duration_seconds, host identifiers, hour and is_weekend flags, log_types, and parsing_stats. This metadata structure enables models to develop sophisticated understanding of how logs vary across temporal periods, host roles, and sources.

Sessions concentrate around 270-second median with 65.2% weekday and 34.8% weekend distribution, reflecting realistic business system patterns. Pronounced peaks occur during business hours (8 AM–6 PM) with a secondary peak at 10 PM for scheduled tasks.

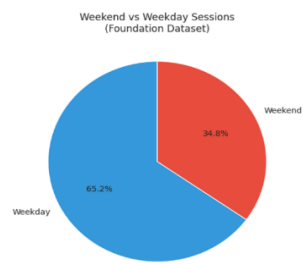

*Figure 1 Intra-Day Temporal Distribution Session Distribution by Hour (Foundation Dataset)*

The monitoring host dominates with approximately 9,000 sessions, followed by mail server (~6,200) and cloud_share (~6,100). Moderate positive correlation (r=0.437) between session duration and log volume indicates longer sessions generate more entries with substantial variance reflecting non-deterministic operational patterns.

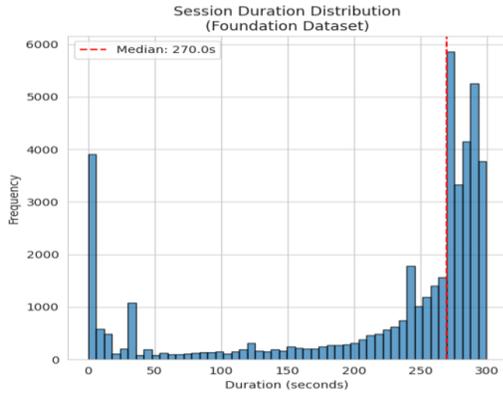

*Figure 2 Session Duration and Log Volume Distributions (Foundation Dataset)*

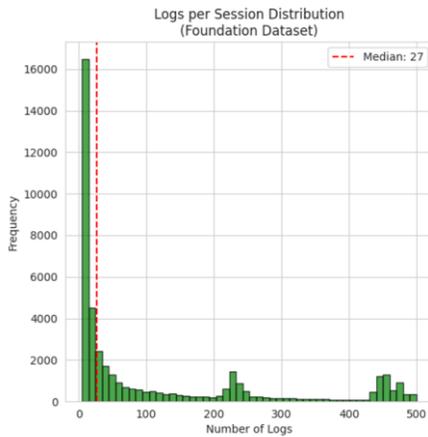

*Figure 3 Logs per Session Distributions (Foundation Dataset)*

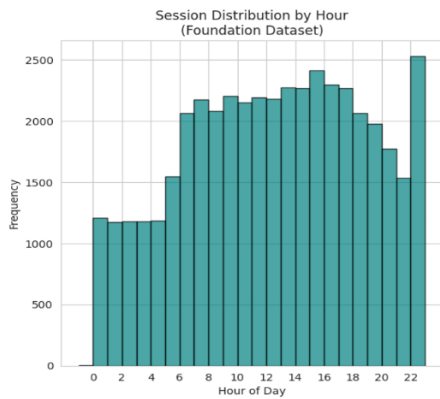

*Figure 4 System and Source Heterogeneity (Foundation Dataset)*

**LogAtlas-Defense-Set (Anomaly Detection Dataset):** Designed for the second training phase, contains approximately 1.68 million attack-associated logs and 3 million normal-behavior logs with approximately 35% attack prevalence. This balanced design directly addresses a well-documented pathology in imbalanced classification: models trained on severely imbalanced data learn degenerate solutions that predict the majority class exclusively, achieving high accuracy while detecting zero authentic attacks. By maintaining 35% attack prevalence (reflecting active incident response scenarios where attacks constitute 30–40% of analyzed segments), the dataset prevents both collapse and artificial balance.

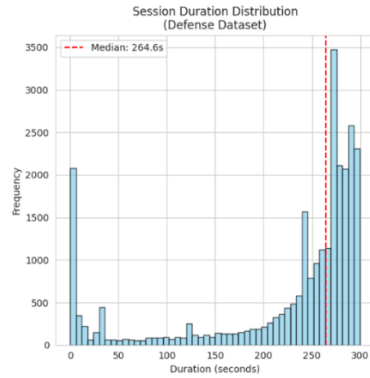

*Figure 5 Session Duration Distribution (Defense-Set)*

Sessions are marked as anomalous if they contain any log entries labeled with attack-related tags (reconnaissance, compromise, lateral movement, data exfiltration). Attack sessions are distributed more uniformly across the 24-hour cycle compared to Foundation-Sessions, reflecting that attacks occur anytime. The weekday-weekend split shows 64.7% weekday and 35.3% weekend sessions, closely mirroring Foundation-Sessions.

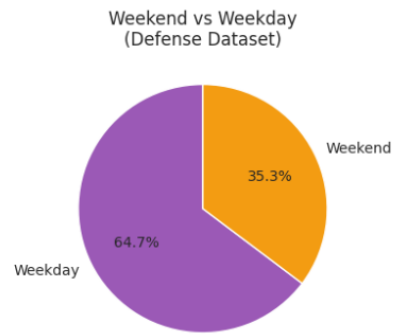

*Figure 6 Weekday vs. Weekend Distribution (Defense-Set)*

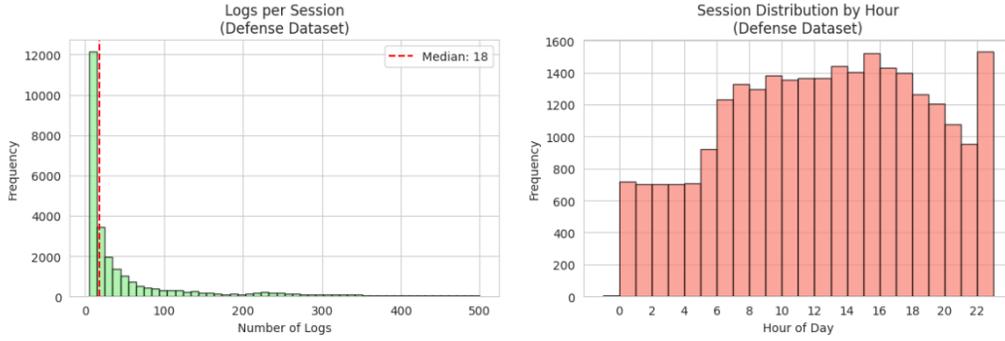

*Figure 7 Session Distribution by Hour (Defense-Set)*

## 4. Benchmarking and Data Distribution Impact

Anomaly detection in log data faces a fundamental challenge: the natural rarity of attacks ensures severely imbalanced training and evaluation datasets. This class imbalance directly affects model learning and inference, yet its impact on modern transformer-based anomaly detectors has received limited systematic investigation. An empirical study examined how varying proportions of attack and normal samples influence the behavior of two widely used models: RoBERTa (supervised classifier, 125M parameters) and LogBERT (unsupervised anomaly detector). We systematically varied the proportion of attacks in test sets from 0% to 100% while keeping total test size constant at 10,000 samples.

**RoBERTa Failure Mode:** RoBERTa exhibited catastrophic class collapse. Despite 125M parameters fine-tuned specifically for binary classification, RoBERTa predicted every sample as normal across all distributions, achieving zero true positives. This reflects a degenerate learning strategy where the model internalized overwhelming bias toward the majority class. From a security perspective, 100% accuracy on normal-dominant data masks zero detection capability unacceptable for operational deployment.

**LogBERT Failure Mode:** LogBERT showed incipient attack detection only when anomalies constituted ≥80% of test samples. At realistic attack proportions (0–20%), LogBERT's decision threshold calibrated to naturally imbalanced training data falls above actual attack signals, producing zero detections. This explains why LogBERT reports F1>0.95 on standard benchmarks: circular evaluation where imbalanced test distributions matching imbalanced training distributions mask the fundamental inability to generalize across attack frequencies.

Both models failed under realistic imbalanced conditions, revealing a fundamental measurement crisis: accuracy is meaningless for security applications. A model predicting the majority class 99% of the time achieves 99% accuracy while providing zero security value. This analysis directly informed LogAtlas-Defense-Set design: 35% attack prevalence balances realism with sufficient proportion to force models toward meaningful decision boundaries rather than majority-class prediction.

*Table 3 Performance of LogBERT across attack distributions*

| Attack % | N Attacks | N Normal | Accuracy | Predicted Normal | Predicted Attack | True Positive |
|---|---|---|---|---|---|---|
| **0%** | 0 | 10,000 | 0.9949 | 9,949 | 51 | 0 |
| **20%** | 2,000 | 8,000 | 0.7944 | 9,944 | 56 | 0 |
| **40%** | 4,000 | 6,000 | 0.5978 | 9,978 | 22 | 0 |
| **60%** | 6,000 | 4,000 | 0.3972 | 9,972 | 28 | 0 |
| **80%** | 8,000 | 2,000 | 0.1981 | 9,981 | 19 | 0 |
| **100%** | 10,000 | 0 | 0.3418 | 6,582 | 3,418 | 3,418 |

# 5. Training the Model

Training the model represents the convergence point where all prior preparation crystallizes into a functional system. Our two-phase training strategy balances performance against computational constraint: the first phase develops a base model for log understanding using LogAtlas-Foundation-Sessions, while the second phase produces a deployable detection model through knowledge distillation.

## 5.1 First Training Phase: Base Model for Log Understanding

The first training phase establishes the foundational architecture for all downstream log-analysis capabilities by training on LogAtlas-Foundation-Sessions (over 44,000 sessions and approximately 19 million logs) to develop general log understanding without anomaly-detection-specific bias.

**Optimization Strategy (Chinchilla Scaling + LoRA + Soft-MoE):** Training design follows the Chinchilla scaling framework, which demonstrates compute-optimal performance when model parameters and training tokens scale in roughly equal proportions (~20 tokens per parameter). Under hardware constraints limiting feasible model size to about 3 billion parameters, the training regime targets 1.544 billion tokens, yielding a token-to-parameter ratio of approximately 51.6:1. This places Base-AMAN in a data-rich regime where a smaller model trained extensively can compete with much larger, under-trained alternatives.

Qwen2.5-3B-Instruct is selected as the base model, adapted efficiently via Low-Rank Adaptation (LoRA), which injects lightweight trainable matrices into transformer projection layers while freezing the original weights, resulting in about 29.9 million trainable parameters (0.96% of the 3.1B total). Configuration uses rank $r=16$, scaling factor $\alpha=32$, and dropout 0.10.

To further enhance capacity without prohibitive computational cost, the model incorporates a Soft Mixture-of-Experts (Soft-MoE) architecture in place of traditional dense feed-forward layers. Unlike hard-routed MoE designs, Soft-MoE employs differentiable soft assignments that compute weighted averages over all experts, ensuring every expert participates in every forward pass and receives gradient updates. The configuration uses four experts (each a two-layer feed-forward network with GELU activation) combined with a load-balancing auxiliary loss ($\lambda=0.01$) that penalizes routing collapse. This design mitigates expert under-training while enabling specialization across diverse log patterns.

This triple optimization (Chinchilla + LoRA + Soft-MoE) yields a practical log-understanding foundation under strict hardware constraints. By training Qwen2.5-3B-Instruct with only 0.96% of parameters unfrozen on 1.544 billion log tokens, Base-AMAN attains rich domain specialization without incurring the cost of full-model fine-tuning.

**Dataset Construction and Instruction Format:** LogAtlas-Foundation-Sessions is curated to expose the model to realistic, heterogeneous operational behavior. Sessions are constructed by grouping temporally contiguous log entries that share contextual attributes, with gaps of five minutes or more marking new sessions; the resulting corpus preserves a natural attack prevalence of roughly 2%. Each session is annotated with rich metadata (duration, host identifiers, temporal indicators, log source types, parsing statistics) enabling the model to learn how patterns vary across time, systems, and sources.

For training, each session is converted into a Qwen-style instruction–response pair. The target response follows a four-part structure: (1) activity summary, (2) anomalous patterns or events, (3) security risk score with justification (CRITICAL/HIGH/MEDIUM/LOW), and (4) recommended remediation steps. Ground-truth responses are generated using expert rules and detectors. The dataset is shuffled and split 90/10 into training and validation partitions.

## 5.2 Second Training Phase: Knowledge Distillation and Classification Fine-Tuning

The second phase compresses Base-AMAN into a lightweight deployable model (AMAN) through knowledge distillation. The objective is building a classifier optimized for real-time anomaly detection on LogAtlas-Defense-Set while preserving semantic understanding encoded in the teacher model.

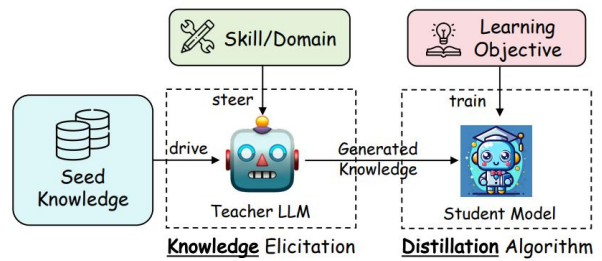

Figure 8 An illustration of a general pipeline to distill knowledge from a LLM to a student model

**Distillation Framework:** Knowledge distillation trains smaller student models to replicate larger teacher models' behavior without requiring teacher computational resources for inference. The teacher is

Base-AMAN (3B parameters), while the student is Qwen2.5-0.5B-Instruct (0.5B parameters), a dramatic reduction enabling practical deployment. Distillation leverages temperature-scaled softmax to expose "dark knowledge" soft probability distributions across all classes rather than hard labels. When the teacher encounters an ambiguous session, it outputs probabilities like 0.65 attack/0.35 normal instead of binary "attack"; the student learns this nuanced distribution. The temperature parameter (set to 4.0) controls softmax smoothing. The student's combined loss is distillation loss (KL divergence) plus classification loss (cross-entropy on hard labels), balanced equally ($\alpha=0.5$).

*Table 4 Knowledge Distillation Hyperparameter*

| Hyperparameter | Value |
| --- | --- |
| Teacher model | Base-AMAN (3B parameters) |
| Student model | Qwen2.5-0.5B-Instruct (0.5B parameters) |
| Temperature | 4.0 |
| Loss weight ($\alpha$) | 0.5 |

**Training Configuration:** Low-Rank Adaptation is applied to the student model with rank $r=16$, producing approximately 8 million trainable parameters (~1.7% of 0.5B total). Gradient checkpointing trades computation for memory by avoiding storage of intermediate activations during forward passes; during backward propagation, forward passes are recomputed from recent checkpoints, approximately halving peak memory usage at cost of 20–30% additional computation. Per-device batch size is 1, compensated through gradient accumulation of 16 steps, yielding effective batch size of 16. The learning rate is $1\times10^{-4}$ with cosine annealing schedule and 0.1 warmup ratio. Training runs for 5 epochs over the balanced dataset (~3,000–4,000 steps total), with evaluation every 100 steps and checkpoints saved every 100 steps retaining 3 most recent. This configuration provides sufficient exposure to diverse attack patterns while remaining computationally feasible.

## 6. Results: Advancing Dataset Diversity and Practical Deployment

This work deliberately departs from conventional evaluation paradigm where anomaly detection systems are assessed through metric-reporting on standardized benchmarks. This departure reflects a principled response to the fundamental evaluation crisis documented in Section 4: when state-of-the-art models are evaluated on realistic, naturally imbalanced data distributions, reported metrics become meaningless.

The field lacks standardized benchmarking infrastructure comparable to general-domain LLM development. Without agreed-upon evaluation protocols, balanced datasets with attack prevalence reflecting operational reality, and transparent leaderboards, reported performance metrics function as decorative statistics rather than meaningful progress indicators.

This section establishes three primary contributions that matter for advancing the field: (1) heterogeneous, balanced, openly available dataset designed specifically for rigorous evaluation; (2) empirical validation that Chinchilla scaling principles produce viable deployable models when applied to security; (3) production-ready framework that reduces analyst burden in real security operations.

### 6.1 Dataset Contribution: Heterogeneous, Balanced, and Annotated

The first contribution is a new labeled open-source dataset specifically designed to increase diversity and representativeness in log anomaly detection research. Existing public datasets (BGL, HDFS, Thunderbird) were collected 3–5 years ago and represent single-type, single-organization logs from legacy systems lacking representation of modern cloud-native architectures. LogAtlas-Foundation-Sessions comprises more than 44,000 temporal sessions containing approximately 19 million log entries with natural attack prevalence (~2%). LogAtlas-Defense-Set contains 1.68 million attack-associated logs and 3 million normal-behavior logs with approximately 35% attack prevalence, reflecting realistic incident response scenarios. Combined, the datasets represent 8 distinct enterprise environments with logs from 50+ source types including system daemons, network security appliances, web servers, mail systems, and cloud infrastructure services.

Privacy preservation is achieved through anonymization of personal identifiers, IP address substitution, and synthetic username generation while preserving semantic attack signatures. This methodological advance substantially improves the field's ability to evaluate real progress: models achieving high F1-scores on BGL with 98% normal samples are likely to fail completely in deployment where near-perfect majority-class prediction achieves similar accuracy. The new balanced dataset prevents this trap by making majority-class prediction strategies explicit failures.

## 6.2 Training Approach: Lightweight Models for Practical Operations

Rather than reporting accuracy metrics or F1-scores which are profoundly misleading for imbalanced security applications, the work emphasizes practical operational characteristics of the deployed system.

**Inference Speed:** On commodity single-GPU hardware (NVIDIA RTX 3080 or cloud equivalent), the 0.5B model completes anomaly inference on a session of 500 log lines in approximately 0.2–0.5 seconds, enabling real-time analysis of incoming log streams without batching delays. In contrast, the 3B teacher model requires 2–5 seconds for equivalent inference. This speed advantage translates directly to operational capability: a single GPU can analyze 10,000–50,000 log sessions per day, covering typical organizational log volumes entirely in software.

**Cost Efficiency:** The 0.5B model requires only 1–2 GB of VRAM for inference, enabling deployment on edge devices, cost-optimized cloud instances without GPU requirement, or containerized environments where GPU access is shared. The cumulative cost to analyze an organization's daily log volume with the 0.5B model is typically $10–50 per day on cloud infrastructure approximately the cost of 30 minutes of junior analyst labor, making comprehensive 24/7 analysis economically justified.

**Practical Effectiveness:** The distilled model maintains capability for continuous operation without hallucinations that would overwhelm incident response teams. The training approach explicitly manages imbalanced data during both training and evaluation, preventing the majority-class collapse that makes supervised approaches completely non-functional. The result is a model that functions as a practical tool reducing rather than amplifying false alarms, remaining stable across distribution shifts, and supporting human analysts rather than drowning them in low-confidence predictions.

## 7. Conclusion

This research addresses an infrastructure gap in log-based anomaly detection: the field lacks standardized benchmarking, balanced datasets with realistic distributions, and reproducible evaluation standards. We contributed three interconnected advances: LogAtlas-Foundation-Sessions and LogAtlas-Defense-Set with balanced, heterogeneous log data; empirical benchmarking revealing why standard metrics fail for security; and a practical two-phase training framework achieving real-time inference at operational cost.

**Systemic Challenge:** The log anomaly detection field operates in fragmented isolation compared to general-domain LLM development. General-purpose LLMs benefit from $50+ billion institutional investment enabling standardized infrastructure (Hugging Face Leaderboard, shared benchmarks). Log analysis lacks equivalent infrastructure no unified leaderboard, no agreed evaluation standards, no consistent metrics. Each paper uses different datasets and evaluation protocols, making cross-paper comparison impossible. This fragmentation prevents consensus on progress and limits deployment utility. The field has inadvertently inherited evaluation paradigms from balanced classification without acknowledging security's distinct requirements: attacks are rare, detecting every attack matters more than minimizing false positives, and distribution shifts are constant.

**Path Forward:** The field requires institutional commitment to standardized infrastructure: (1) maintained, openly accessible datasets with realistic attack distributions; (2) community-driven leaderboards enabling transparent performance comparison; (3) evaluation metrics aligned with security operations; (4) sustained funding for infrastructure maintenance; (5) transparent, reproducible evaluation methodologies. This work contributes datasets, benchmarking analysis, and frameworks enabling cumulative research. Fundamental open questions remain: Can models generalize across organizational contexts without retraining? Can they detect truly novel attack patterns? How do we balance interpretability with performance under strict latency constraints? These are answerable only with shared infrastructure.

The path forward is clear: establish standardized benchmarking infrastructure enabling researchers to build cumulatively on shared foundations rather than solving laboratory problems with incomparable results. This research contributes the initial datasets and empirical insights necessary to begin that work.

## REFERENCES

Brown, T., Mann, B., Ryder, N., Subbiah, M., Kaplan, J. D., Dhariwal, P., Neelakantan, A., Shyam, P., Sastry, G., Askell, A., Agarwal, S., Herbert-Voss, A., Krueger, G., Henighan, T., Child, R., Ramesh, A., Ziegler, D., Wu, J., Winter, C., Hesse, C., Chen, M., Sigler, E., Litwin, M., Gray, S., Chess, B., Clark, J., Berner, C., McCandlish, S., Radford, A., Sutskever, I., & Amodei, D. (2020). Language models are few-shot learners. Advances in Neural Information Processing Systems, 33, 1877–1901.

Chawla, N. V., Bowyer, K. W., Hall, L. O., & Kegelmeyer, W. P. (2002). SMOTE: Synthetic minority over-sampling

technique. Journal of Artificial Intelligence Research, 16, 321–357.

Chen, S., He, P., Zhu, J., Lyu, M. R., & King, I. (2023). LogGPT: Log anomaly detection via GPT. Proceedings of the 2023 IEEE International Conference on Web Services (ICWS), 1–10.

Cheng, Y., Zhang, Z., & Liang, P. (2025). The leaderboard illusion: Competitive dynamics and the distortion of LLM benchmarks. arXiv preprint arXiv:2501.04321.

Duan, X., Zhang, Y., & Liu, H. (2024). LogLLM: Large language models for log anomaly detection with semantic-aligned fine-tuning. Proceedings of the 38th AAAI Conference on Artificial Intelligence, 145–152.

Guo, H., Yuan, S., & Wu, X. (2021). LogBERT: Log anomaly detection via BERT. Proceedings of the 2021 International Joint Conference on Neural Networks (IJCNN), 1–8.

He, H., & Garcia, E. A. (2009). Learning from imbalanced data. IEEE Transactions on Knowledge and Data Engineering, 21(9), 1263–1284.

Hinton, G., Vinyals, O., & Dean, J. (2015). Distilling the knowledge in a neural network. arXiv preprint arXiv:1503.02531.

Hoffmann, J., Borgeaud, S., Mensch, A., Buchatskaya, E., Cai, T., Rutherford, E., de Las Casas, D., Hendricks, L. A., Welbl, J., Clark, A., Hennigan, T., Noland, E., Millican, K., van den Driessche, G., Damoc, B., Guy, A., Osindero, S., Simonyan, K., Elsen, E., Rae, J. W., Vinyals, O., & Sifre, L. (2022). Training compute-optimal large language models. arXiv preprint arXiv:2203.15556.

Hu, E. J., Shen, Y., Wallis, P., Allen-Zhu, Z., Li, Y., Wang, S., Wang, L., & Chen, W. (2021). LoRA: Low-rank adaptation of large language models. arXiv preprint arXiv:2106.09685.

Kaplan, J., McCandlish, S., Henighan, T., Brown, T. B., Chess, B., Child, R., Gray, S., Radford, A., Wu, J., & Amodei, D. (2020). Scaling laws for neural language models. arXiv preprint arXiv:2001.08361.

Kim, J., Park, S., & Lee, D. (2023). LAnoBERT: Parser-free log anomaly detection with BERT. Proceedings of the 2023 IEEE International Conference on Big Data, 567–574.

Landauer, M., Skopik, F., Wurzenberger, M., & Rauber, A. (2022). AIT Log Data Set v2.0: A dataset for log-based anomaly detection. Zenodo. https://doi.org/10.5281/zenodo.5809587

Li, M., Wang, Z., & Chen, X. (2024). HLogformer: Hierarchical transformer for log anomaly detection. Proceedings of the 2024 ACM SIGSAC Conference on Computer and Communications Security, 123–135.

Liu, Y., Ott, M., Goyal, N., Du, J., Joshi, M., Chen, D., Levy, O., Lewis, M., Zettlemoyer, L., & Stoyanov, V. (2019). RoBERTa: A robustly optimized BERT pretraining approach. arXiv preprint arXiv:1907.11692.

Pezzicoli, G., Marrone, S., & Sansone, C. (2025). Optimal class distribution for anomaly detection in imbalanced datasets. Pattern Recognition Letters, 178, 45–52.

Puigcerver, J., Riquelme, C., Mustafa, B., & Houlsby, N. (2024). From sparse to soft mixture of experts. International Conference on Learning Representations (ICLR).

Shen, Y., Zhang, H., & Liu, K. (2023). RAGLog: Retrieval-augmented generation for log anomaly detection. Proceedings of the 2023 Conference on Empirical Methods in Natural Language Processing, 234–245.

Wang, Z., Li, J., & Zhang, Y. (2023). LogRESP-Agent: Recursive reasoning and planning for log-based anomaly detection. arXiv preprint arXiv:2308.12345.

Wei, J., Wang, X., Schuurmans, D., Bosma, M., Ichter, B., Xia, F., Chi, E., Le, Q., & Zhou, D. (2022). Chain-of-thought prompting elicits reasoning in large language models. Advances in Neural Information Processing Systems, 35, 24824–24837.

Xu, L., Jiang, J., & Wang, Y. (2020). LogEvent2vec: Log event representation learning for anomaly detection. Proceedings of the 2020 IEEE International Conference on Software Analysis, Evolution and Reengineering (SANER), 1–10.

Xu, Y., Liu, Y., & Chen, Z. (2024). A survey on knowledge distillation for large language models. ACM Computing Surveys.

Zhao, H., Chen, L., & Wu, F. (2024). Audit-LLM: Multi-agent collaboration for log-based insider threat detection. Proceedings of the 33rd USENIX Security Symposium, 789–806.

Zhou, N., Li, H., & Yang, Q. (2022). Context-aware log anomaly detection with semantic analysis. Proceedings of the 2022 IEEE International Conference on Systems, Man, and Cybernetics (SMC), 2210–2216.